\documentclass[modern, a4paper, times]{aastex63}
\usepackage{amsmath}
\usepackage{graphicx}
\usepackage{natbib}

\submitjournal{ApJ}

\shorttitle{Solar UV Anisotropy Effects on Helioglow}
\shortauthors{Strumik et al.}

\newcommand{\kms}{~km~s$^{-1}$}

\begin{document}

\title{Effects of Heliolatitudinal Anisotropy of Solar FUV/EUV Emissions on Lyman-$\alpha$ Helioglow}

\correspondingauthor{M. Strumik}
\email{maro@cbk.waw.pl}

\author[0000-0003-3484-2970]{M. Strumik}
\affil{Space Research Centre PAS (CBK PAN), Bartycka 18a, 00-716 Warsaw, Poland}

\author[0000-0003-3957-2359]{M. Bzowski}
\affil{Space Research Centre PAS (CBK PAN), Bartycka 18a, 00-716 Warsaw, Poland}

\author[0000-0002-5204-9645]{M. A. Kubiak}
\affil{Space Research Centre PAS (CBK PAN), Bartycka 18a, 00-716 Warsaw, Poland}

\begin{abstract}
We present a study of the influence of solar UV anisotropy on the heliospheric backscatter helioglow generated by resonant scattering of solar Lyman-$\alpha$ photons on interstellar hydrogen atoms around the Sun. Simulations based on the WawHelioGlow model suggest that the response of the helioglow pole-to-ecliptic ratio to the anisotropy is linear, but 15\% of the anisotropy (polar darkening) generates 30--40\% change in the ratio in the solar minimum and 15--20\% in the solar maximum. We attribute this difference to an interplay between the solar UV anisotropy and the latitudinal structure of the solar wind in solar minima. The solar UV anisotropy also increases the helioglow intensity from the downwind direction by 5--10\%, due to the influence of the anisotropy on the ionization losses and trajectories of atoms passing by the Sun in polar regions. Consequently, mid-latitude regions (in the heliographic and ecliptic coordinates) are least affected by the UV anisotropy.

By comparison of the simulation results with observations of the SOHO/SWAN satellite instrument, we derive the day-by-day time evolution of the solar Lyman-$\alpha$ anisotropy for the north and south poles over two solar cycles from 1996 to 2022. The inferred anisotropy is $\sim$5--10\% in solar minima and $\sim$15--25\% in solar maxima, the northern anisotropy being stronger than the southern. Our study suggests that in solar minima a highly structured solar wind is associated with relatively small solar UV anisotropy, while in solar maxima the solar wind is more isotropic but a substantial solar UV anisotropy appears.
\end{abstract}

%%%%%%%%%%%%%%%%%%%%%%%%%
\section{Introduction}
\label{sec:intro}
\noindent
The Sun is moving through the interstellar medium with a velocity of $\sim$26 \kms{} towards the ecliptic longitude of $\sim$256$\degr$  and the latitude of $\sim$5$\degr$ \citep{lallement_etal:04a, witte:04, bzowski_etal:08a, bzowski_etal:15a, schwadron_etal:15a, swaczyna_etal:23c, bzowski_etal:23b}. The interaction of the solar wind and the local interstellar matter leads to the formation of the heliosphere, bounded by the heliopause, which is a separatrix for the heliospheric and interstellar plasma flows \citep{baranov:90a, ratkiewicz_etal:98a, izmodenov:00, pogorelov_etal:04a}. The neutral atoms penetrate this structure easily and may reach the vicinity of the Sun. Due to the charge exchange with the solar wind protons and photoionization processes, a cavity around the Sun is formed, where the density of interstellar neutral hydrogen is very low in comparison with that in the unperturbed interstellar medium \citep{bzowski_etal:01a}.

The heliospheric backscatter Lyman-$\alpha$ glow (hereafter helioglow) is an emission generated by the resonant scattering of solar Lyman-$\alpha$ photons on neutral hydrogen of interstellar origin around the Sun \citep{bertaux_blamont:71, thomas_krassa:71}. The helioglow intensity depends on the illuminating flux from the Sun and the distribution of neutral atoms in its proximity. The density and radial velocity of the atoms play an important role in shaping the helioglow-flux distribution in the sky \citep{brasken_kyrola:98}.

Generally, the distribution of the helioglow intensity in the sky shows a dipole-like pattern, with the minimum and maximum in the proximity of the hydrogen downwind and upwind directions, respectively \citep[see, e.g.,][]{lallement_etal:05a}. This specific pattern is related to the shape of the hydrogen cavity around the Sun, which is organized  around the upwind-downwind axis and extended in the downwind direction. However, the position of an observer relative to the upwind-downwind axis and dependencies of the illumination, ionization rates, and radiation pressure on time and heliolatitude introduce some additional modulations to the simple dipole pattern \citep{bzowski:03, quemerais_etal:06b, kubiak_etal:21b}. These additional modulations are a subject of interest in satellite missions (e.g., SOHO/SWAN, \citet{bertaux_etal:95}, or the upcoming IMAP/GLOWS mission, \citet{mccomas_etal:18b}), because they are a potential source of information about the solar wind structure. The possibility of inferring the solar wind structure from the helioglow observations has been presented in several papers \citep{bertaux_etal:95,bzowski:03,bzowski_etal:03a, lallement_etal:10b,koutroumpa_etal:19a, katushkina_etal:19a}, but the obtained results are at some tension with those obtained from analyses of interplanetary scintillations \citep{sokol_etal:13a, sokol_etal:15a, sokol_etal:20a, porowski_etal:22a, porowski_etal:23a}, as demonstrated by \citet{katushkina_etal:13a}.

In this paper, we focus on a relatively poorly understood question of the influence of the anisotropy of the solar UV output on the helioglow distribution in the sky \citep{cook_etal:81a, pryor_etal:92, auchere_etal:05a}. The solar UV anisotropy discussed in this paper can be caused by the presence of polar coronal holes, but also by active regions on the Sun. The distribution of these regions on the Sun's surface depends on both the solar longitude and latitude. The longitudinal modulation was found to be responsible for the so-called searchlight effect in the helioglow on the Carrington-rotation time scale \citep[see][]{bertaux_etal:00, quemerais_bertaux:02b}. In this paper, we consider the UV anisotropy as averaged over the solar longitudes (or Carrington-rotation time scale in observations), thus we assume that it varies only with the heliolatitude.

The solar Lyman-$\alpha$ is one of the main factors responsible for the existence of the helioglow %emission If the solar UV radiation is of concern, the largest influence on the helioglow intensity is by the Lyman-$\alpha$ illumination 
\citep{kubiak_etal:21b}. The wavelength-integrated solar Lyman-$\alpha$ flux is measured on a regular basis, but as an average over the solar disc seen from the Earth orbit \citep{machol_etal:19a}. This type of measurement 
%is affected mostly by the solar equatorial flux and it 
does not allow to estimate the heliolatitudinal structure of the Lyman-$\alpha$ flux emitted away from the Earth's direction, in particular towards the solar polar regions. Indirect methods of inferring the anisotropy include the analysis of images of the Sun in 30.4 nm and correlations between solar emission in the 30.4 nm and Lyman-$\alpha$ wavebands \citep{auchere:05}. However, this type of correlation exhibits a relatively wide statistical spread, which affects the accuracy of the inference. A connection between He 1086 nm images of the Sun and the anisotropy in the solar Lyman-$\alpha$ flux was also studied as an explanation of some latitudinal-anisotropy effects in the interplanetary Lyman-$\alpha$ \citep{pryor_etal:96}. Full-sky images of the Sun in the Lyman-$\alpha$ waveband have very recently started to be provided by the ASO-S/LST instrument \citep{li_etal:19a}. However, to study relations between the helioglow time evolution and solar activity in the past, we need to know the solar Lyman-$\alpha$ anisotropy over an extended period. For this reason, in this paper, we study an indirect method of inferring the time evolution of the solar UV anisotropy from a comparison of the helioglow observations and simulations.

This paper is generally intended as an extension of our previous work \citet{strumik_etal:21b}, where a similar method of inferring the solar UV anisotropy was discussed. In the present work, we clarify the main differences between the effects of the solar-wind and solar-UV anisotropies on the helioglow distribution in the sky. We discuss quantitative estimates of the effects of the solar UV anisotropy in different regions of the sky and different phases of the solar cycle. This allows us to identify regions in the sky that are least affected by the anisotropy effects, which can be of interest if a model with isotropic solar UV is compared with observations. We also investigate temporal effects in the anisotropy-inferring method proposed by \citet{strumik_etal:21b}, which finally leads to a more accurate estimation of the inferred anisotropy provided in the present work.

In Section \ref{sec:modeling}, we discuss briefly the WawHelioGlow model used for simulations and its settings for the study presented in this paper. We also discuss the results of simulations of the influence of the solar UV anisotropy on the helioglow both for all-sky maps and for lightcurves representing specific cuts through the sky maps. In Section \ref{sec:model_vs_obs}, we compare the simulation results with observations from the SOHO/SWAN instrument. Based on this comparison, we infer the time evolution of the solar UV anisotropy during the past two solar cycles (1996--2022). Our findings are summarized in Section \ref{sec:discussion}.

\section{Insights from Modeling}
\label{sec:modeling}
\noindent
In this section, we use the WawHelioGlow model of the helioglow \citep{kubiak_etal:21a} to investigate the effects of the solar UV anisotropy on the backscatter Lyman-$\alpha$ helioglow as seen by an observer located at $\sim$1 au from the Sun.
We start by briefly discussing the model and its settings, then present analyses of all-sky maps. 
With this, we discuss how the effects of solar-wind and solar-UV anisotropies manifest in selected simulated lightcurves of the helioglow for the viewing geometry similar to that planned for the forthcoming IMAP/GLOWS experiment.

\subsection{Model description}
\noindent
The WawHelioGlow simulation code calculates the intensity of the helioglow for a selected epoch, vantage point, and viewing direction \citep{kubiak_etal:21a}. The model is implemented in a modular way, where a more or less realistic input can be used for simulations of the helioglow. The helioglow intensity is computed from the first principles, but the model can perform data-driven simulations if a realistic time series are provided as the model input. Optically thin single-scattering approximation is used to compute the heliglow intensity by integration of a source function along the lines of sight. The source function in a given point in space is assumed to depend on the solar irradiance and the distribution function of hydrogen atoms, mainly the density and radial velocity.
	
In the WawHelioGlow model, the distribution of the hydrogen around the Sun is computed from the (n)WTPM model \citep{tarnopolski_bzowski:09}, where the temporal and heliolatitudinal modulation of ionization losses can be taken into account. The effects of the charge exchange, photoionization, and electron-impact ionization are included. The radiation pressure is computed in a self-consistent way as the momentum change resulting from the solar Lyman-$\alpha$ flux accounting also for the Doppler-shift effects dependent on the radial velocity of the hydrogen atoms.

\subsubsection{Solar UV anisotropy}
In the WawHelioGlow model, the irradiances for the illuminating (Lyman-$\alpha$) and photoionizing fluxes can be set as dependent on the heliolatitude $\phi$ in the following way
\begin{equation}
E=E_\mathrm{eq} \left(a \sin^2\phi +\cos^2\phi\right)
\label{eq:anis}
\end{equation}
(see Section 2.6.3 in \citet{kubiak_etal:21a} for details). The anisotropy parameter can be defined as $a=E_\mathrm{p}/E_\mathrm{eq}$, where $E_\mathrm{p}$ and $E_\mathrm{eq}$ are the polar and equatorial irradiances, correspondingly. As shown by \citet{kubiak_etal:21a}, after simple algebraic transformations Equation (\ref{eq:anis}) is equivalent to the anisotropy definition used by \citet{pryor_etal:92}. The anisotropy parameter for the simulations discussed in this section is set to either $a=0.85$ or $a=1$. The value of $a=0.85$ corresponds to 15\% of solar-longitude-averaged polar darkening of the solar UV output. One should bear in mind that the anisotropy depends on the distance of the observer from the Sun as discussed by \citet{auchere_etal:05a} and \citet{strumik_etal:21b}. The WawHelioGlow setting of $a=0.85$ should be understood as the effective value for hydrogen atoms from regions significantly contributing to the helioglow, which are located far ($\gtrsim$1 au) from the Sun. The distant-observer polar darkening of 15\% corresponds to $\sim$50\% of the anisotropy close to the solar surface according to a simple model presented by \citet{strumik_etal:21b}. In the current WawHelioGlow implementation, the anisotropy defined by Equation (\ref{eq:anis}) is constant in time and symmetric relative to the heliographic equator, thus the same for both poles. Results presented by \citet{strumik_etal:21b} and further in this paper suggest that generally different anisotropy levels are expected for the north and south poles and the anisotropy is expected to evolve in time. These capabilities are planned to be included in future versions of the WawHelioGlow model, but the possibility of using these features requires prior knowledge of a realistic time evolution of the anisotropy for both poles separately, which could be set as the model input. Such a phenomenological model of the solar UV anisotropy is derived in this paper.

\subsubsection{Model input and simulation settings}
In its most-realistic-input setting, the WawHelioGlow model uses the charge-exchange rates from \citet{porowski_etal:22a}, the photoionization rates from \citet{sokol_etal:20a}, the radiation pressure from \citet{IKL:20a} (based on the Lyman-$\alpha$ flux from \citet{machol_etal:19a}), and the electron-impact ionization from \citet{bzowski_etal:12b} (based on the OMNI2 solar wind data discussed in \citet{king_papitashvili:05}).

\begin{figure}[!htbp]
	\begin{center}
		\includegraphics[scale=0.6]{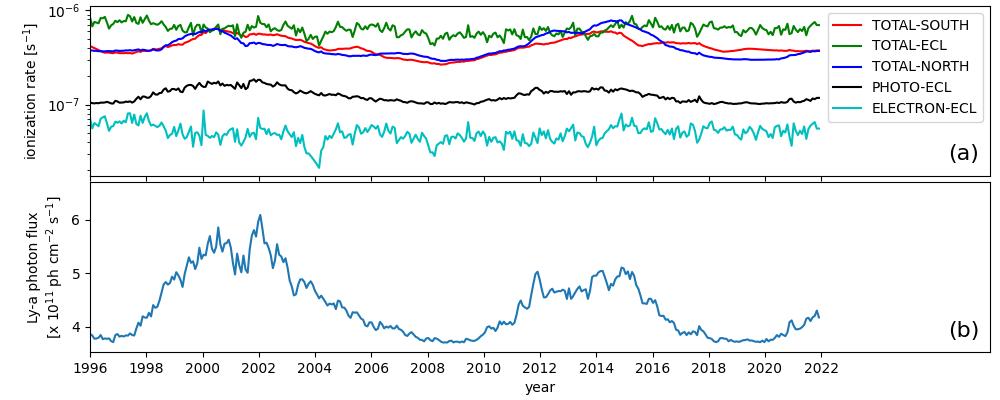}
		\caption{Time dependence of (a) ionization rates and (b) Lyman-$\alpha$ photon flux at 1 au from the Sun from 1996 to 2022. Three total ionization rates (north -- dark blue, ecliptic -- green, and south -- red) are compared in panel (a) with the photoionization (black) and electron-impact ionization (cyan) ecliptic rates. The ecliptic Lyman-$\alpha$ flux in panel (b) is based on \citet{machol_etal:19a}.}
		\label{fig:ionization_and_lya_vs_time}
	\end{center}
\end{figure}
Figure \ref{fig:ionization_and_lya_vs_time} illustrates the time dependence of the ionization rates and solar Lyman-$\alpha$ photon flux at 1 au from the Sun. The total ionization rate at the poles (red and dark blue lines) is comparable to the ecliptic rate (green) in solar maxima of $\sim$2001 and $\sim$2014. In solar minima ($\sim$1996, $\sim$2010, and $\sim$2019), significant decreases in the polar rates are observed, which suggests a substantial total-ionization-rate anisotropy. The ionization rates affect the helioglow by changing the distribution of the hydrogen around the Sun. The solar-cycle-related decreases of the total polar rates seen in Figure \ref{fig:ionization_and_lya_vs_time}(a) can be expected to change quasi-periodically the shape of the hydrogen cavity around the Sun, moving the boundary of the cavity towards the Sun at high heliographic latitudes near solar minima. Since the solar Lyman-$\alpha$ irradiance changes approximately like $\propto\! r^{-2}$ with the heliodistance $r$, such deformation of the cavity increases the helioglow backscattered flux from polar regions.

The ecliptic photoionization (black) and electron-impact ionization (cyan) rates shown in Figure \ref{fig:ionization_and_lya_vs_time}(a) are small in comparison with the total rate. The dominant contribution to the total rate is the charge-exchange reaction (see, e.g., \cite{sokol_etal:20a}), which depends on the density and velocity of the solar wind. The heliolatitude-dependent solar wind speed used here is derived from ground-based observations of interplanetary scintillations (IPS) as presented by \citet{porowski_etal:22a}. This phenomenological model of the solar wind parameters uses also the OMNI data from the ecliptic plane and latitudinally-invariant solar wind energy flux for calculating the solar wind density. The model accounts for the solar cycle variations of the solar wind, but it is important to note that helioglow observations (e.g., from the SOHO/SWAN instrument) are not used in any way in deriving the solar wind structure.

The bottom panel shows the ecliptic Lyman-$\alpha$ flux based on \citet{machol_etal:19a}, which also exhibits a solar-cycle modulation. The Lyman-$\alpha$ flux affects the helioglow in two ways, by changing the illumination of the hydrogen atoms and the radiation pressure. The illumination effect is the dominant one for the helioglow intensity in comparison with the effect of changes in the radiation pressure, which affects the trajectories of the hydrogen atoms and thus indirectly also the helioglow.

\begin{figure}[!htbp]
	\begin{center}
		\includegraphics[scale=0.51]{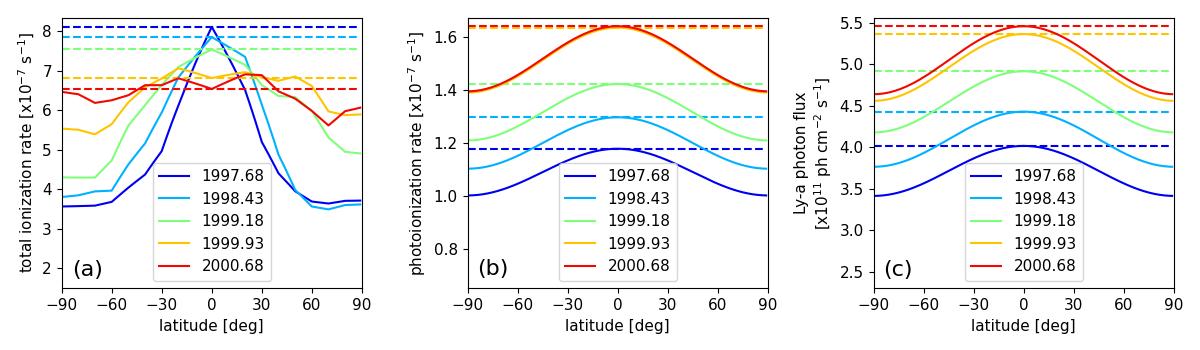}
		\caption{Heliolatitudinal dependence of (a) total ionization rate, (b) photoionization rate, (c) Lyman-$\alpha$ photon flux at 1 au from the Sun. Different colors correspond to five selected dates to illustrate the evolution of the quantities between the solar minimum and maximum. Solid lines correspond to latitudinally-modulated quantities (referred to as 3D in the text) and dashed lines show isotropic case (2D in the text). For the photoionization (panel (b)) and Lyman-$\alpha$ flux (panel (c)),  the anisotropy level in Equation (\ref{eq:anis}) is assumed to be $a=0.85$.}
		\label{fig:ionization_and_lya_vs_lat}
	\end{center}
\end{figure}

In Figure \ref{fig:ionization_and_lya_vs_lat} we show heliolatitudinal profiles of the total ionization rate, photoionization rate, and Lyman-$\alpha$ photon flux for some selected dates to illustrate in more detail how the profiles change from solar-minimum conditions ($\sim$1997.68) to solar-maximum conditions ($\sim$2000.68).

We use three types of model settings in this paper. In the first setting, hereafter referred to as $\mathrm{SW2D/UV2D}$, the heliolatitudinal dependencies of both the solar wind and solar UV output are neglected and the ecliptic-equatorial parameters of the solar wind and irradiance are used for all heliolatitudes. In particular, it means that the rates of both of the ionization reactions (charge-exchange and photoionization) depend only on time and radial distance from the Sun, which can be considered as a two-dimensional model of the variability of the rates (hence the 2D label). This type of setting corresponds to dashed lines in Figure \ref{fig:ionization_and_lya_vs_lat}. Note that in this approach, the modeled radiation pressure depends on time and heliodistance, but one should bear in mind also the dependence on the radial component of the velocity of hydrogen atoms moving throughout the heliosphere.

The second setting, hereafter $\mathrm{SW3D/UV2D}$, is extended relative to the former by including only the heliolatitudinal structure of the solar wind and consequently the charge-exchange ionization rate. In terms of the relations illustrated in Figure \ref{fig:ionization_and_lya_vs_lat}, it would be equivalent to a model using the solid-line dependencies from panel (a) and dashed-line relationships from panels (b) and (c). In the third setting, henceforth $\mathrm{SW3D/UV3D}$, the full capabilities of the model are unleashed, and both the solar wind and solar UV output (both illuminating Lyman-$\alpha$ and photoionizing flux) are modeled as dependent on time, heliodistance, and heliolatitude (hence 3D label). Solid-line relations from all panels of Figure \ref{fig:ionization_and_lya_vs_lat} are used in the third case.

For the simulations of $\mathrm{SW3D}$-type, the WawHelioGlow model uses a state-of-the-art realistic time-dependent and heliolatitude-dependent solar wind structure discussed by \citet{porowski_etal:22a}. Reducing the model to $\mathrm{SW2D}$-type means that only OMNI-database ecliptic-equatorial measurements are used and the solar wind is assumed to be isotropic. For the solar UV radiation, the time dependence is taken from observations \citep{sokol_etal:20a, IKL:20a, machol_etal:19a}, but the anisotropy is handled in a simplified manner defined in Equation (\ref{eq:anis}). For the simulations of $\mathrm{UV2D}$-type, i.e., $\mathrm{SW2D/UV2D}$ and $\mathrm{SW3D/UV2D}$, the anisotropy parameter is set to $a=1$, and for $\mathrm{SW3D/UV3D}$ we set $a=0.85$.

In simulations presented in this paper, exactly the same parameter values for the interstellar hydrogen inflow are used as in \citet{kubiak_etal:21b}, which are explicitly listed in Table 1 in \citet{IKL:18b}.

\subsection{Simulation results -- comparison of all-sky maps}
\label{sec:maps_cmp}
\begin{figure}[!htbp]
	\begin{center}
		\includegraphics[scale=0.28]{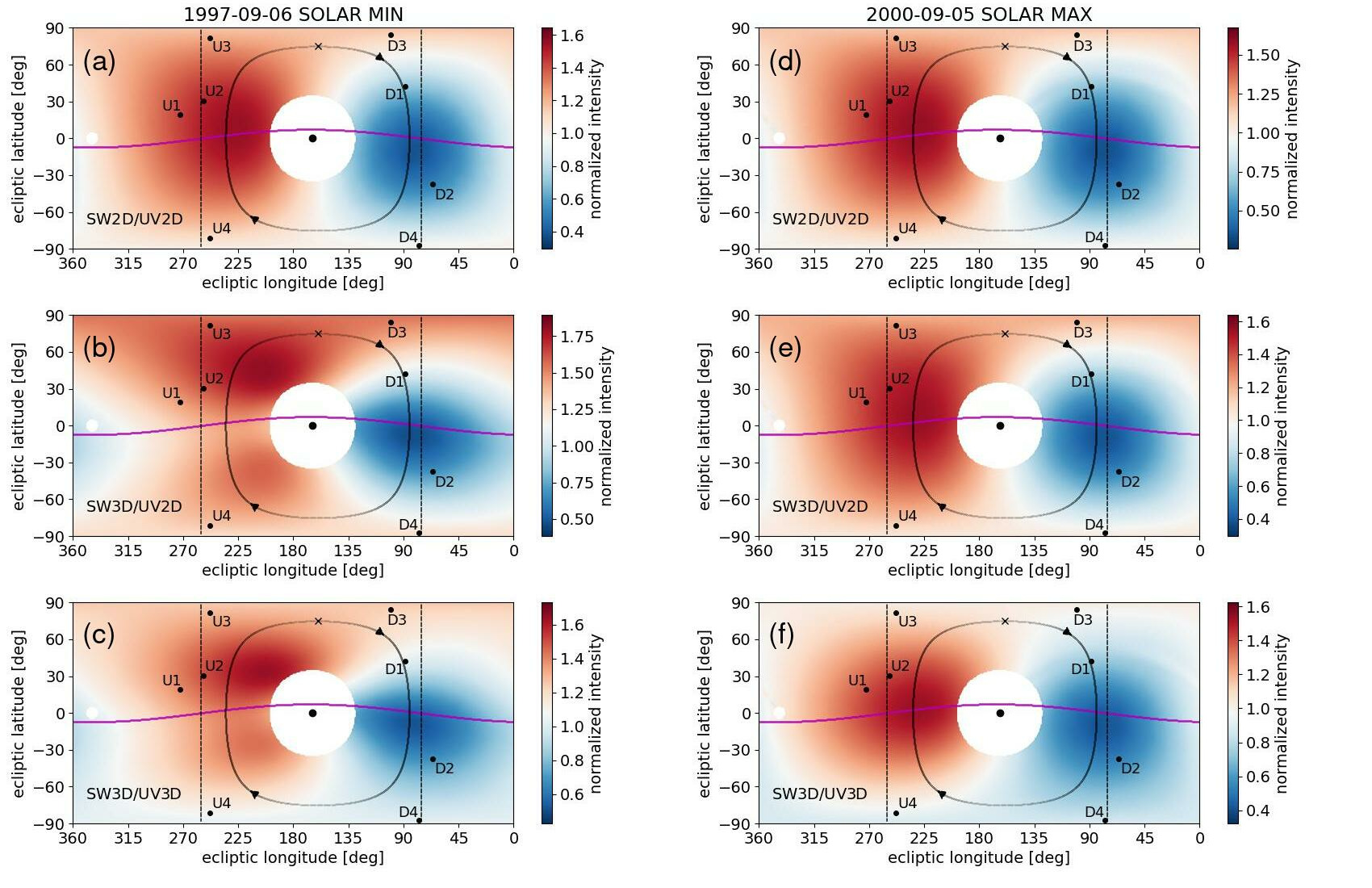}
		\caption{Comparison of sky maps of the normalized intensity obtained in WawHelioGlow simulations for three different models of the solar wind and solar UV structure. Model name is specified by a label in the bottom left part of each map. The difference between 2D and 3D models consists in including the heliolatitudinal dependence (3D) in addition to the time and heliodistance dependence (2D) for the solar wind (SW) and solar UV output (UV). Selected dates correspond to the crosswind position of SOHO satellite (located at $\sim$1 au from the Sun). Left column (a)-(c) shows differences for the solar minimum conditions (Sep 6, 1997) and the right column (d)-(f) -- for the solar maximum (Sep 5, 2000). The ecliptic coordinates are used and the inverted-axis convention for the horizontal axis (longitude) to make the maps corresponding to the real sky. The magenta line shows the heliographic equator. The black-line circle represents a scanning circle of the angular radius of 75\degr{} with its center $\sim$4\degr{} off the solar direction. The arrows on the circle show the direction of increasing spin angle and the cross shows the zero-spin position (the northernmost point of the circle). Black vertical lines are constant-longitude cuts through the map at the downwind (longitude of $75\fdg 4$) and upwind (longitude of $255\fdg 4$) directions. The selected directions D1-D4 (U1-U4) close to the downwind (upwind) directions in the sky are used for analyses presented in the paper. The white circular region was excluded from computations due to its closeness to the Sun (black small circle).}
		\label{fig:maps}
	\end{center}
\end{figure}

\noindent
Figure \ref{fig:maps} shows
simulated sky maps of the normalized helioglow intensity for the three models mentioned above. The simulations were done for solar-minimum and solar-maximum conditions for an observer located in the ecliptic plane at a crosswind point at 1 au. The influence of the heliolatitudinal structure of the solar wind can be apprehended by the comparison of panels (a) and (b), and the influence of the solar UV anisotropy -- by the comparison of panels (b) vs. (c) for the solar minimum, and (e) vs. (f) for the solar maximum.

\begin{figure}[!htbp]
	\begin{center}
		\includegraphics[scale=0.28]{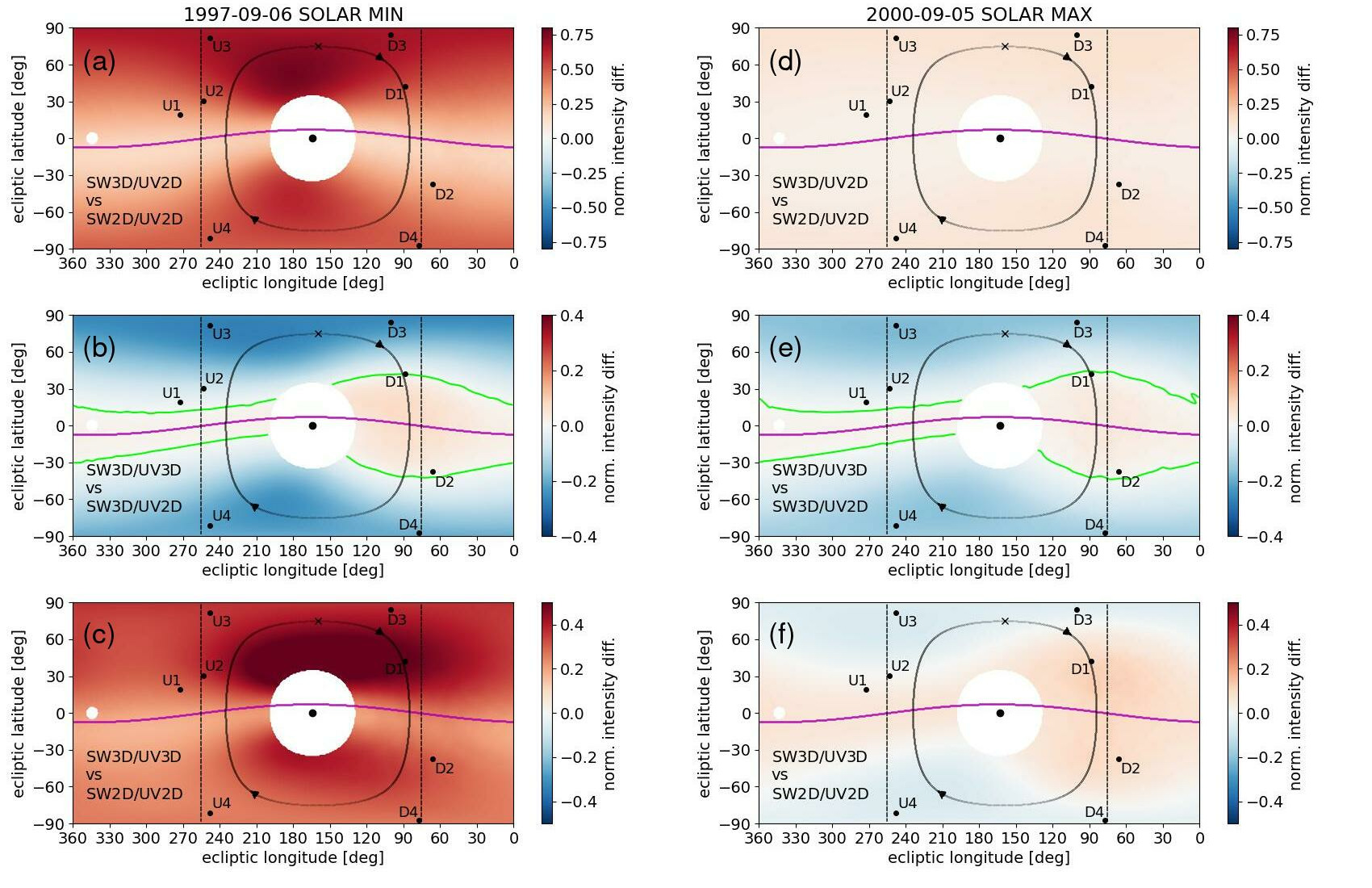}
		\caption{Normalized intensity differences [see Equation (\ref{eq:DeltaI})] between the three models shown in Figure \ref{fig:maps}. Model-pair name is specified by a label in the bottom left part of each map. The green lines show the zero-value isocontour for the intensity difference. Additional elements (lines, labels, etc.) and conventions used in the maps are described in the caption of Figure \ref{fig:maps}.}
		\label{fig:cmp_maps}
	\end{center}
\end{figure}

To make the comparisons easier, Figure \ref{fig:cmp_maps} presents differences between the three types of simulations shown in Figure \ref{fig:maps}.
All maps in Figure \ref{fig:cmp_maps} show a normalized intensity difference
\begin{equation}
\Delta I = (I_\mathrm{M1}-I_\mathrm{M2})/\langle I_\mathrm{EQ} \rangle,
\label{eq:DeltaI}
\end{equation}
where $I_\mathrm{M1}$ and $I_\mathrm{M2}$ are the intensities computed from models M1 and M2, respectively. The normalization intensity is computed as $\langle I_\mathrm{EQ} \rangle = (\langle I_\mathrm{M1} \rangle + \langle I_\mathrm{M2} \rangle)/2$, where $\langle I_\mathrm{M1} \rangle$, $\langle I_\mathrm{M2} \rangle$ are averaged values over the $\pm 10\degr$ belt around the ecliptic equator for a given pair of models, M1 and M2.
 
In Figure \ref{fig:cmp_maps} (a) and (d), a difference between the models $\mathrm{SW3D/UV2D}$ (anisotropic solar wind, isotropic solar UV) and $\mathrm{SW2D/UV2D}$ (isotropic solar wind and UV) is presented. In the solar minimum (panel (a)), the polar regions in the helioglow become brighter than the equatorial, which illustrates the influence of the heliolatitudinal structure of the solar wind on the helioglow. This structure typically leads to a lower charge-exchange ionization rate in the polar regions, which changes the shape of the hydrogen cavity, moving the boundary of the cavity towards the Sun at high heliographic latitudes. In the solar maximum (panel (d)), the solar-wind heliolatitudinal profile is relatively flat (compare the blue and red line in Figure \ref{fig:ionization_and_lya_vs_lat}), and thus no substantial difference between the polar regions and the equator is observed.

Figure \ref{fig:cmp_maps} (b) and (e) illustrates a difference between the models $\mathrm{SW3D/UV3D}$ (anisotropic solar wind and solar UV) and $\mathrm{SW3D/UV2D}$ (anisotropic solar wind, isotropic solar UV), thus related to effects introduced by the solar UV anisotropy in a model with a realistic structure of the solar wind. A striking (but naturally expected) feature is that the polar darkening works in the opposite direction than the solar-wind heliolatitudinal structure, i.e., the UV anisotropy causes the polar regions to be significantly darker, both in the solar minimum and maximum. One can also see an increased signal from the equatorial region, mainly from the downwind direction (longitude of $\sim$$75\fdg4$).
This is consistent with expectations because some atoms from the downwind region passed  the Sun in high-latitude regions and were subjected to smaller photoionization losses and radiation pressure as compared to the conditions of isotropic solar UV output. This leads to increased hydrogen density and helioglow intensity in the downwind direction. Consequently, the regions least affected by the UV anisotropy (close to the green lines) are located at a certain angular distance from the heliographic equator (magenta line) at mid latitudes, especially in the downwind sector.

In Figure \ref{fig:cmp_maps} (c) and (f), a difference between simulations $\mathrm{SW3D/UV3D}$ (anisotropic solar wind and solar UV) and $\mathrm{SW2D/UV2D}$ (isotropic solar wind and solar UV) is presented. For the solar minimum (panel (c)), the combined effects of anisotropies of the solar wind and solar UV output are visible here, where the maximum difference between the two models occurs in mid-latitude regions. For the solar maximum (panel (f)), the difference between the models is similar to panel (e), because the lack of significant solar wind anisotropy makes the UV anisotropy the main factor shaping the distribution of the difference in the sky.

\subsection{Simulation results -- cuts through the all-sky maps}
\label{sec:cuts_cmp}
\noindent
The maps shown in the previous section are convenient for a global qualitative description of the solar-wind and solar-UV anisotropy effects on the helioglow, but for quantitative analysis, it is useful to define some cuts through the maps. The first type of cut is along a scanning circle shown in all maps in Figure \ref{fig:cmp_maps}. The circle has an angular radius of 75\degr{} and its center is shifted by $\sim$4$\degr${} relative to the Sun's position in the sky, which represents the planned observation geometry of the GLOWS instrument onboard the upcoming IMAP mission \citep{mccomas_etal:18b}. Another type of analyzed cut is for constant longitudes of 75\fdg 4 (close to downwind) and 255\fdg 4 (upwind), which are also shown in Figure \ref{fig:cmp_maps}.

\begin{figure}[!htbp]
	\begin{center}
		\includegraphics[scale=0.4]{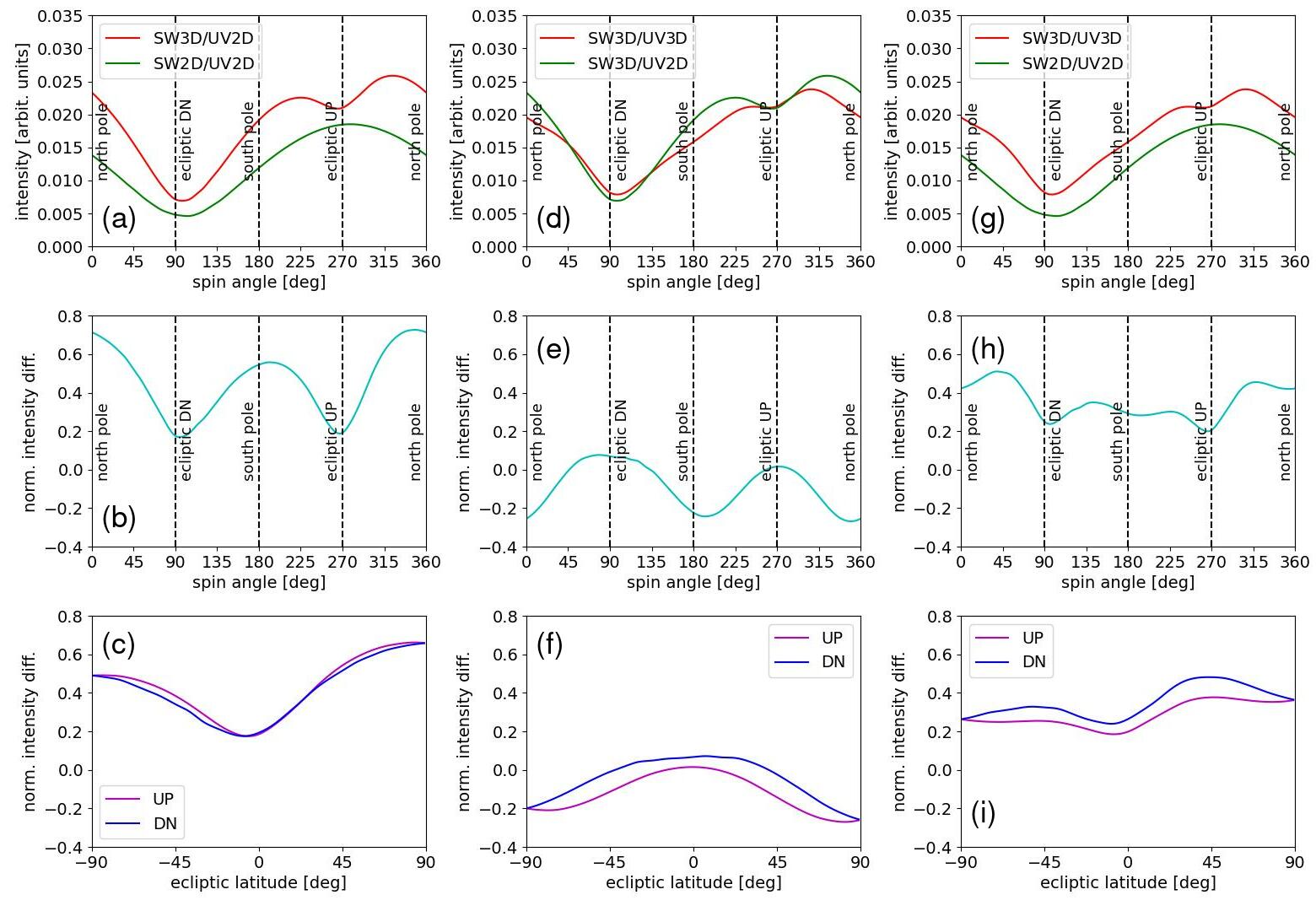}
		\caption{Comparison of scanning-circle (the first and second row) and constant-longitude (the third row) cuts through the all-sky maps shown in Figure \ref{fig:cmp_maps}. The geometry of the cuts is explained in the caption of Figure \ref{fig:cmp_maps}. Helioglow-intensity differences between three types of model settings are considered here: (a)--(c) $\mathrm{SW3D/UV2D}$ vs. $\mathrm{SW2D/UV2D}$, (d)--(f) $\mathrm{SW3D/UV3D}$ vs. $\mathrm{SW3D/UV2D}$, and (g)--(i) $\mathrm{SW3D/UV3D}$ vs. $\mathrm{SW2D/UV2D}$ for the solar minimum conditions on September 6, 1997.}
		\label{fig:cmp_sol_min}
	\end{center}
\end{figure}

Figure \ref{fig:cmp_sol_min} shows a comparison of different pairs of models for the solar-minimum conditions.
Differences related to the anisotropic solar wind structure ($\mathrm{SW3D/UV2D}$ vs. $\mathrm{SW2D/UV2D}$) are shown in panels (a)--(c). In the absence of the solar wind anisotropy, a typical helioglow modulation along the GLOWS-like scanning circle (Figure \ref{fig:cmp_sol_min}(a) -- green line) is relatively simple with its minimum in the downwind segment and maximum in the upwind part. Solar wind anisotropy introduces an additional modulation (red line) for this particular vantage point and increases the average intensity of the helioglow. These effects are better seen in panel (b), where the normalized difference of the intensities suggests that the brightening is from 20\% in the ecliptic region to up to 70--80\% close to the poles. This is confirmed by the latitudinal dependence of the difference shown in panel (c). 

The differences caused by the solar UV anisotropy ($\mathrm{SW3D/UV3D}$ vs. $\mathrm{SW3D/UV2D}$) are presented in Figure \ref{fig:cmp_sol_min} (d)--(f). Panels (d) and (e) show that the solar UV anisotropy on average decreases the intensity of the helioglow and introduces a modulation anticorrelated with the effects due to the solar wind anisotropy. As seen in panel (e), for the solar UV anisotropy of 15\% set in the model, the helioglow darkening may be up to 30\% for the polar directions. However, in the proximity of the ecliptic plane in the downwind direction, a $\sim$10\% brightening of the helioglow is observed, which was explained in the previous section. At the upwind direction close to the ecliptic, a very small ($\sim$1\%) intensity increase is seen. Note, that the polar darkening of 30\% mentioned above can be interpreted as a change of the pole-to-ecliptic ratio related to a change of the solar UV anisotropy by 15\%.

Figure \ref{fig:cmp_sol_min} (g)--(i) shows the combined effects of strong solar wind and solar UV anisotropies ($\mathrm{SW3D/UV3D}$ vs. $\mathrm{SW2D/UV2D}$). At first sight, the helioglow intensity (red line) shown in panels (a) and (g) is similar. However, the extraction of modulations that are on top of the general upwind-downwind dipole pattern is conveniently done when the pattern due to isotropic solar wind and solar UV is subtracted. This approach has been applied in panels (b), (e), and (h), where the combined effects of the solar wind and UV anisotropies seem to lead to an intermediate state (panel (h)) between the two states shown in panels (b) and (e). For a strong solar wind anisotropy, the addition of the UV anisotropy leads to a decrease of the average intensity, reduces the amplitude of the intensity modulation, and introduces an additional modulation due to the anticorrelation between the solar-wind and solar-UV anisotropy effects. In particular, the addition of the UV anisotropy may lead to mid-latitude bumps in the normalized intensity difference as clearly seen in panel (i).

\begin{figure}[!htbp]
	\begin{center}
		\includegraphics[scale=0.4]{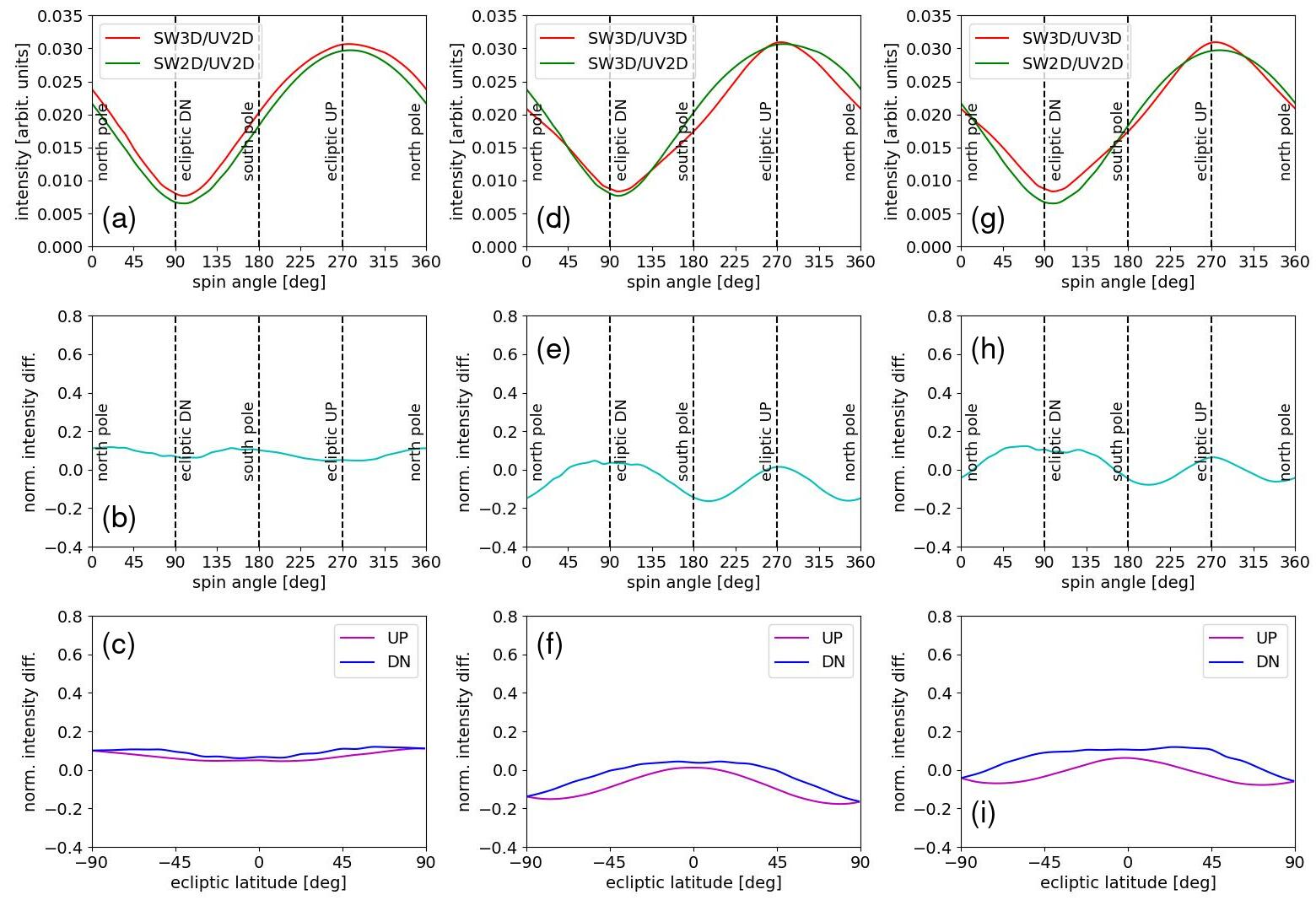}
		\caption{Similar comparison as in Figure \ref{fig:cmp_sol_min}, but for the solar maximum conditions on Sep 5, 2000.}
		\label{fig:cmp_sol_max}
	\end{center}
\end{figure}

It is well known that during solar maxima, variations of the solar wind density and velocity with heliolatitude are rather weak \citep[see, e.g.,][]{porowski_etal:23a}. A case of relatively flat solar wind structure (Sep 5, 2000; corresponding to the red line labeled ``2000.68'' in Figure \ref{fig:ionization_and_lya_vs_lat}) is presented in Figure \ref{fig:cmp_sol_max}, where the effects of the solar UV anisotropy on the helioglow in the presence of this weakly modulated solar wind structure are illustrated. Panels (a)--(c) ($\mathrm{SW3D/UV2D}$ vs. $\mathrm{SW2D/UV2D}$)
show a difference of the intensity of $\sim$10\%, with a very weak modulation along the scanning circle if only the effects of the solar wind structure are concerned. Adding a solar UV anisotropy of 15\% ($\mathrm{SW3D/UV3D}$ vs. $\mathrm{SW3D/UV2D}$) results in  variations with a higher amplitude, as seen in panels (d)--(f), but the decrease of the helioglow intensity seems to be  15--20\% (panel (e)), i.e., weaker than during the solar minimum (cf. Figure \ref{fig:cmp_sol_min}(e)). Again, on average the solar wind anisotropy and the solar UV anisotropy work in opposite directions. The combined effects of the solar UV anisotropy and weakly-modulated solar wind ($\mathrm{SW3D/UV3D}$ vs. $\mathrm{SW2D/UV2D}$) are extracted in panels (g)--(i), which suggests that the variability of the helioglow difference (relative to $\mathrm{SW2D/UV2D}$ model) during the solar maximum is dominated by the influence of the solar UV anisotropy. However, on average the helioglow intensity changes only slightly (see Figure \ref{fig:cmp_sol_max}(h)), in contrast to the solar minimum case, where it may increase by 20--40\% (see Figure \ref{fig:cmp_sol_min}(h)).

\subsection{Pole-to-ecliptic ratio vs. solar UV anisotropy}
\begin{figure}[!htbp]
	\begin{center}
		\includegraphics[scale=0.7]{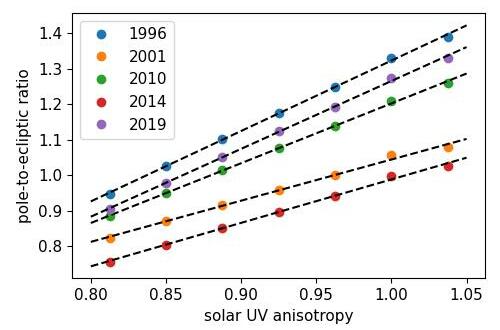}
		\caption{Dependence of the polar-to-ecliptic ratio for the helioglow on the solar UV anisotropy. Simulated effects for times close to solar minima (1996, 2010, 2019) and close to solar maxima (2001, 2014) are illustrated here for an observer located upwind (in our case the position of the SOHO satellite on $\sim$June 5 of each year) at 1 au. Dashed lines show linear fits to the simulated dependencies (circles) for each year.}
		\label{fig:linear_dependence}
	\end{center}
\end{figure}

\noindent
A comparison of panel (e) in Figures \ref{fig:cmp_sol_min} and \ref{fig:cmp_sol_max} suggests that the same level of solar UV anisotropy influences the pole-to-ecliptic intensity ratio for the helioglow to a different extent depending on the solar-cycle phase. This is consistent with Figure 3 in \citet{strumik_etal:21b}, where a bunch of lines showing the pole-to-ecliptic ratio for different anisotropy levels is squeezed in solar maxima as compared with the solar-minima intervals. Figure \ref{fig:linear_dependence} shows how the 
ratio depends on the anisotropy for some selected cases analyzed by \citet{strumik_etal:21b} (upwind-observer geometry). In solar minima, 15\% of solar UV anisotropy leads to even 30-40\% change in the polar-to-ecliptic ratio for the helioglow, while in the solar maxima, the anisotropy of 15\% changes the ratio by only 15-20\%. The ratio is elevated for the solar-minimum conditions (1996, 2010, 2019) relative to the solar maxima (2001, 2014), but the dependence remains linear for both cases as shown in Figure \ref{fig:linear_dependence}. The linear dependence suggests that this different response of the helioglow to the solar UV anisotropy is related to a simple process, where the effects of the solar wind structure (i.e., the average intensity increase and brigther-pole type of modulation seen in Figure \ref{fig:cmp_sol_min}(b)) undergo a secondary modulation by the solar UV anisotropy effects, which finally leads to a significant amplification of the latter during the solar minima. During the solar maxima, a much weaker modulation due to the solar wind structure causes a weaker amplification effect of the solar UV anisotropy on the pole-to-ecliptic intensity ratio.

\section{Solar FUV/EUV anisotropy retrieved from SOHO/SWAN observations}
\label{sec:model_vs_obs}
\noindent
The previous section shows that the anisotropy of the solar UV output may be an important factor affecting the distribution of the helioglow intensity in the sky. In this section, we investigate day-by-day evolution of the solar UV anisotropy in the Lyman-$\alpha$ wavelength based on a direct comparison of satellite observations and simulation results. Accurate simulations of the helioglow distribution in the sky are computationally expensive%, because they require computations of the hydrogen distribution in the heliosphere as affected by time, heliodistance, and heliolatitude-varying solar wind structure and anisotropic solar UV output
. To shorten the computation time, we carefully selected a set of eight directions in the sky for which WawHelioGlow simulations were performed to obtain the day-by-day evolution of the helioglow intensity. 
\begin{table}
\caption{Ecliptic coordinates of the directions selected for analysis}
	\centering
	\begin{tabular}{|c|c|c|}
		\hline
		Label & ecl. longitude & ecl. latitude \\ \hline
		D1 & $88\fdg 5$ & $42\fdg 5$ \\ \hline
		D2 & $66\fdg 5$ & $-37\fdg 5$ \\ \hline
		D3 & $100\fdg 5$ & $84\fdg 5$ \\ \hline
		D4 & $77\fdg 5$ & $-87\fdg 5$ \\ \hline
		U1 & $273\fdg 5$ & $20\fdg 5$ \\ \hline
		U2 & $253\fdg 5$ & $30\fdg 5$ \\ \hline
		U3 & $248\fdg 5$ & $81\fdg 5$ \\ \hline
		U4 & $248\fdg 5$ & $-81\fdg 5$ \\ \hline
	\end{tabular}
\label{tab:dirs_dn_up}
\end{table}
Besides the vicinity of the directions to the upwind and downwind longitudes, the main criteria for the selection included also the requirement that half of the selected points is in the polar regions and the other half is at mid latitudes, closer to the ecliptic equator. Moreover, the selection process involved also the examination of the SOHO/SWAN maps for the presence of point sources, to finally select directions possibly least affected by extra-heliospheric contributions.
The selected directions, denoted by D1--D4 (close to the downwind longitude) and U1--U4 (close to the upwind longitude), are shown in Figure \ref{fig:cmp_maps}, and their coordinates are listed in Table \ref{tab:dirs_dn_up}.

A list of dates of SOHO/SWAN observations used in this paper is restricted with respect to all available observation days. Several criteria were used for excluding particular days. By using these criteria, a vast majority of spikes in the time series of the helioglow intensity evolution were removed as illustrated in Figure \ref{fig:cleaning}. A more technical discussion of the data cleaning process can be found in Appendix \ref{sec:appendix1}.

The absolute calibration and its corrections for the sensitivity decrease are difficult problems for satellite instruments performing measurements in the UV range (see, e.g., \citet{quemerais_etal:13} or \citet{frey_etal:17a}). As discussed recently by \citet{baliukin_etal:19a} and \citet{koutroumpa_etal:19a}, the SOHO/SWAN instrument showed a significant sensitivity degradation from 2008 to 2014, which can be expected to change its calibration over time. Due to these calibration uncertainties, we decided to use normalized quantities for our comparisons of the simulated and observed fluxes. Technical details of the normalization procedure are discussed in Appendix \ref{sec:appendix2}.

\subsection{SOHO/SWAN vs. WawHelioGlow -- mid-latitude regions}
\begin{figure}[!htbp]
	\begin{center}
		\includegraphics[scale=0.6]{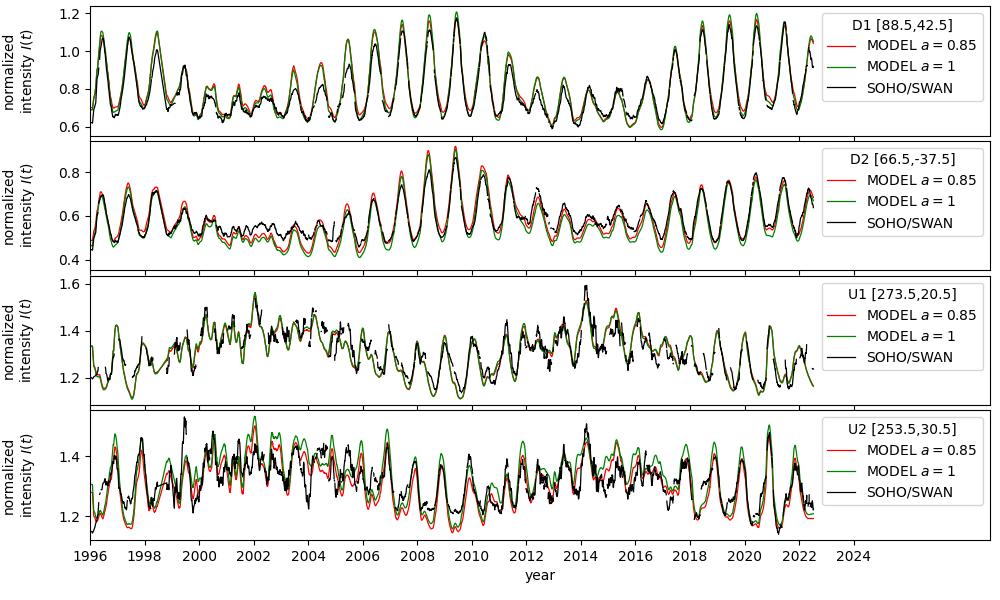}
		\caption{Comparison of WawHelioGlow simulations (red lines for the anisotropy $a=0.85$ and green lines for $a=1$) and SOHO/SWAN observations (black lines). Two top panels show the time evolution of the normalized intensity for the downwind sector (directions D1 and D2), and the bottom two panels -- for the upwind sector (directions U1 and U2). Generally, all the directions are from the mid-latitude part of the sky, where the solar UV anisotropy has little influence on the helioglow.}
		\label{fig:model_obs_cmp}
	\end{center}
\end{figure}

\noindent
Figure \ref{fig:model_obs_cmp} shows a comparison of the normalized modeled and observed
intensities for the directions D1, D2, U1, and U2. The intensities are computed as 30-day moving averages to suppress variations at the Carrington-rotation time scale, which are usually present in the SOHO/SWAN observations but not in the results of the WawHelioGlow model, where Carrington-period-averaged ionization and illumination models are used. For consistency, the averaging procedure was applied to both observations and simulations. In Figure \ref{fig:model_obs_cmp}, one can see oscillations of the intensity on the one-year time scale, which is a well-known phenomenon for the helioglow observed by satellites operating at the Earth orbit, related to periodic changes of the vantage point. For the downwind sector (two top panels) the effects of the solar cycle are seen mainly as a modulation of the amplitude of the yearly oscillations. Smaller-amplitude one-year oscillations and a solar-cycle oscillation are clearly seen for the upwind sector (two bottom panels). %However, one should bear in mind that solar-cycle modulations are partially filtered out from Figure \ref{fig:model_obs_cmp}, because they are effectively included in the normalization factor $I^\mathrm{raw}_0(t)$.
Solar cycle modulations persist even though they are partially included in the normalization factor $I^\mathrm{raw}_0(t)$ defined in Appendix \ref{sec:appendix2}.

The comparison of the model and observations in Figure \ref{fig:model_obs_cmp} shows a general agreement between the model and observations for the selected directions. Small differences between the simulations with different anisotropy settings (red and green lines) also confirm that conclusions from Sections \ref{sec:maps_cmp} and \ref{sec:cuts_cmp} regarding the small influence of the solar UV anisotropy on mid-latitude helioglow intensity are valid not only for the crosswind vantage point but for all vantage points. One should bear in mind that the solar-wind heliolatitudinal structure, and consequently the charge-exchange ionization rates used in the simulations presented in Figure \ref{fig:model_obs_cmp} are derived from interplanetary scintillations \citep{tokumaru_etal:10a, tokumaru_etal:12b, sokol_etal:20a, porowski_etal:22a}, which is a source of information entirely independent of the SOHO/SWAN observations. By using this observed solar-wind heliolatitudinal structure, the WawHelioGlow model becomes a data-driven model, but the agreement between the model and observations presented in Figure \ref{fig:model_obs_cmp} does not result from any fitting procedure, but is obtained as a result of computations from first principles. This suggests that results of WawHelioGlow simulations can be reconciled with SOHO/SWAN observations when time-dependent parameters of the anisotropy of the solar UV output are identified and implemented in the code.

\subsection{Inferring the solar UV anisotropy from helioglow observations}
\begin{figure}[!htbp]
	\begin{center}
		\includegraphics[scale=0.6]{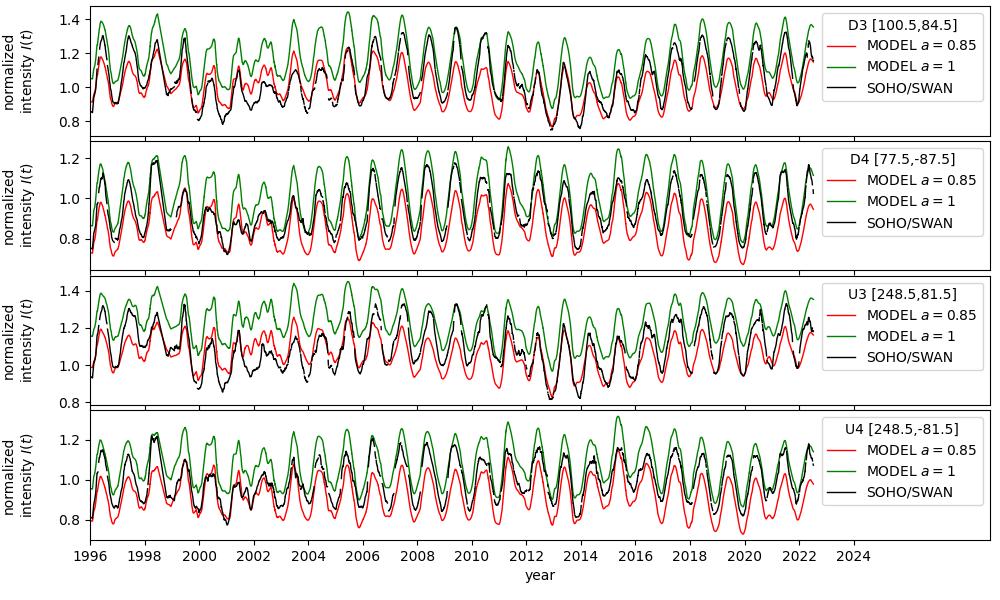}
		\caption{Comparison of the normalized intensities simulated by the WawHelioGlow model ($a=0.85$ -- red line, $a=1$ -- green line) and observed by the SOHO/SWAN instrument (black line). Four directions D3, D4, U3, U4 (from the top) in the polar sectors are shown, where the influence of the solar UV anisotropy is significant.}
		\label{fig:anis_simple_pre}
	\end{center}
\end{figure}

As illustrated in Figure \ref{fig:anis_simple_pre} for the polar directions D3, D4,
U3, and U4, the solar UV anisotropy of the magnitude of 15\% generates significant differences between the red and green lines in the normalized intensity. The SOHO/SWAN observations (black line) are generally contained between the two simulated curves for $a=0.85$ and $a=1$. Therefore, it is possible to use these time series to infer the day-by-day evolution of the solar UV anisotropy $A$ in a similar way as proposed by \citet{strumik_etal:21b}, using the relation
\begin{equation}
A=0.85+0.15~\frac{I_\mathrm{obs}-I_\mathrm{a=0.85}}{I_\mathrm{a=1}-I_\mathrm{a=0.85}},
\label{eq:anis_infer}
\end{equation}
where $I_\mathrm{obs}$ is the normalized intensity observed by SOHO/SWAN (black line in Figure \ref{fig:anis_simple_pre}), and $I_\mathrm{a=0.85}$ ($I_\mathrm{a=1}$) is the modeled intensity for $a=0.85$ ($a=1$), represented by the red (green) line in Figure \ref{fig:anis_simple_pre}. Equation (\ref{eq:anis_infer}) is based on an assumption of a linear correlation between the anisotropy level and polar intensity, which was demonstrated by \citet{strumik_etal:21b} and is fully supported by Figure \ref{fig:linear_dependence}.

\begin{figure}[!htbp]
	\begin{center}
		\includegraphics[scale=0.6]{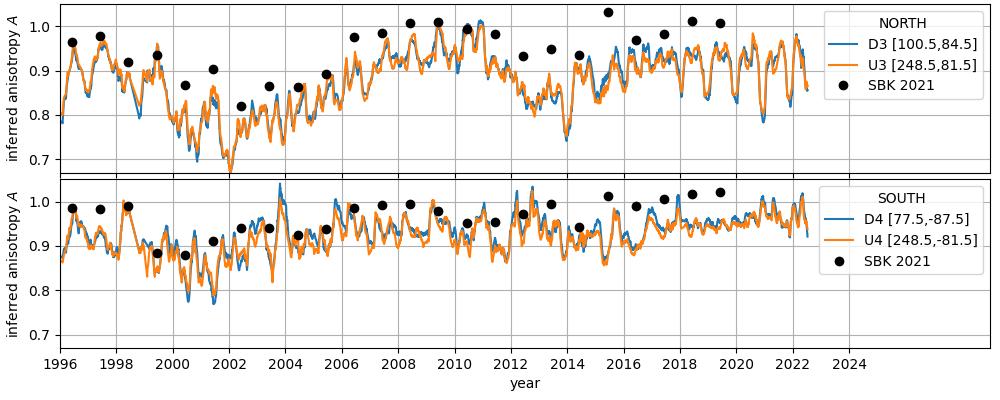}
		\caption{The solar UV anisotropy (blue and orange lines) inferred from the time series presented in Figure \ref{fig:anis_simple_pre} and Equation (\ref{eq:anis_infer}). The top panel shows the results for the north-pole sector, and the bottom panel - for the south pole. Results obtained by \citet{strumik_etal:21b} are shown by black circles labeled ``SBK 2021''.}
		\label{fig:anis_simple_infer}
	\end{center}
\end{figure}

The anisotropy inferred from the time series in Figure \ref{fig:anis_simple_pre} (using Equation (\ref{eq:anis_infer})) is shown in Figure \ref{fig:anis_simple_infer}.
The estimates for different directions, represented by the blue and orange lines in Figure \ref{fig:anis_simple_infer}, are very similar, which is consistent with the expectations because the angular distance in the sky between the points D3 and U3 (also between D4 and U4) is relatively small (but projection effects make the distance apparently larger). The inferred anisotropy evolution (lines) is similar to that presented by \citet{strumik_etal:21b} (shown by black circles labeled ``SBK 2021''). During the solar minima, a smaller anisotropy (i.e., $A$ value closer to 1) is observed as compared to the times of solar maxima, where $A \simeq 0.7-0.8$. The northern anisotropy is generally larger than the southern. The anisotropies are also different during the two following solar maxima, being larger for $\sim$2001 than for $\sim$2014.

However, a detailed inspection of Figure \ref{fig:anis_simple_infer} reveals some periods when a significant annual oscillation of the inferred anisotropy is observed (e.g., 2018-2022 for the northern anisotropy). Such periodicity is unlikely to occur for the solar UV anisotropy. A more likely explanation for the annual modulation is a worse agreement between the simulations and observations for some periods. Indeed, if we take a look at the first and third panels in Figure \ref{fig:anis_simple_pre} for the time intervals 1996-2000 and 2018-2022, we can see that the amplitude of oscillations for the SOHO/SWAN observations (black line) is larger than that obtained in the simulations (red and green lines). This results in observations being closer to the anisotropic simulations (where $a=0.85$) for the downwind position (December of each year) of the satellite, while for other seasons the observations are generally closer to the isotropic simulations ($a=1$). 

\begin{figure}[!htbp]
	\begin{center}
		\includegraphics[scale=0.6]{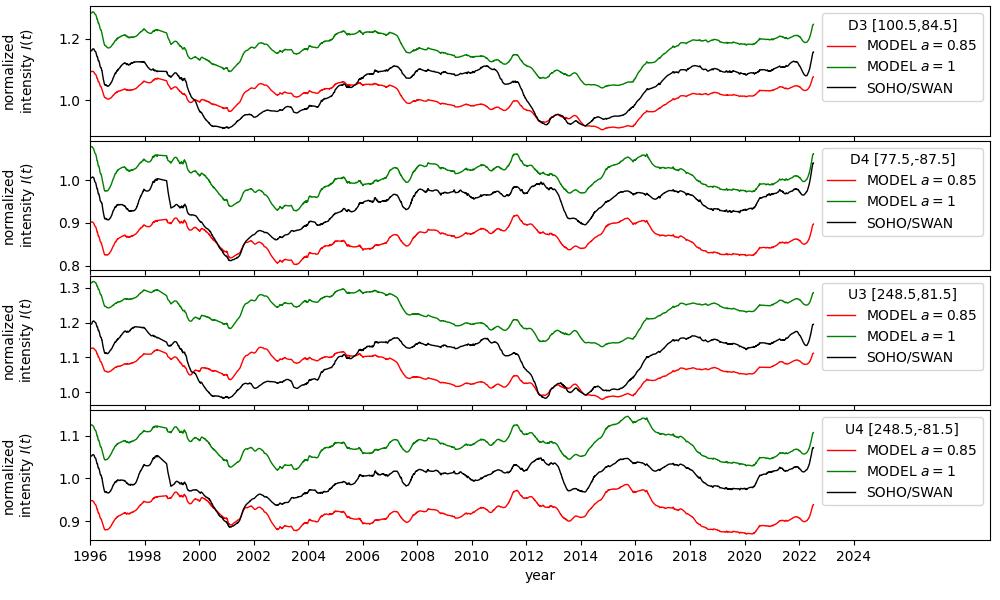}
		\caption{A similar comparison of the normalized simulated and observed intensities as in Figure \ref{fig:anis_simple_pre} but for yearly-moving-average-based time series.}
		\label{fig:anis_yearly_avg_pre}
	\end{center}
\end{figure}

\begin{figure}[!htbp]
	\begin{center}
		\includegraphics[scale=0.6]{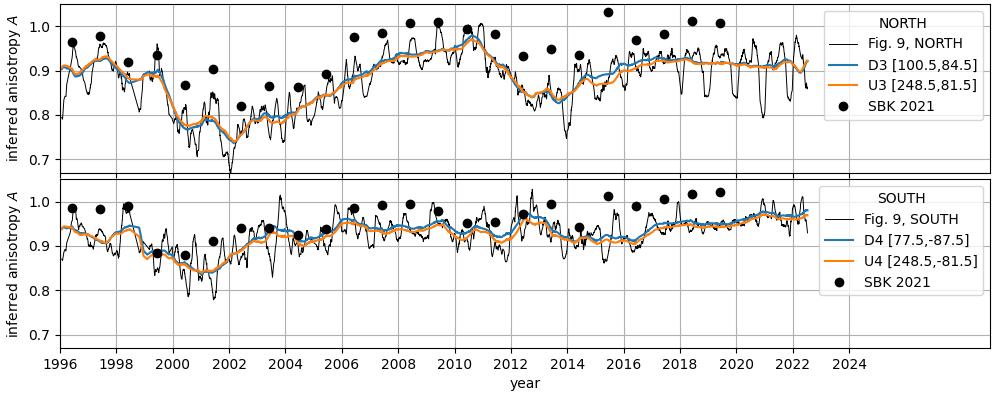}
		\caption{The solar UV anisotropy (blue and orange lines) inferred from Equation (\ref{eq:anis_infer}) and the yearly-moving-average-based time series presented in Figure \ref{fig:anis_yearly_avg_pre}. The top panel shows the results for the north-pole sector, and the bottom panel -- for the south pole. Results obtained by \citet{strumik_etal:21b} are shown by the black circles labeled ``SBK 2021'', and the inferred anisotropy from Figure \ref{fig:anis_simple_infer} (not-averaged case) is shown by the black line.}
		\label{fig:anis_yearly_avg_infer}
	\end{center}
\end{figure}

For this reason, we decided to use a moving average over one year to eliminate the yearly oscillations before the procedure of inferring the solar UV anisotropy. The averaged time series of the normalized intensity are shown in Figure \ref{fig:anis_yearly_avg_pre}.
Except for using the averaged time series, exactly the same procedure as for Figure \ref{fig:anis_simple_infer} is applied then to infer the anisotropy $A$. The results are shown in Figure \ref{fig:anis_yearly_avg_infer},
which shows that the annual modulations are naturally suppressed. Most of the time the yearly-moving-average-based inferred anisotropy (blue and orange lines) is lower than our previous estimate (black circles labeled ``SBK 2021'') based on the upwind observer position. We discuss these results in the next section.

%%%%%%%%%%%%%%%%%%%%%%%%%
\section{Discussion and Conclusions}
\label{sec:discussion}

In this paper, we studied the influence of the solar UV anisotropy on the helioglow distribution in the sky as seen by an observer at 1 au.  We show that an anisotropy of $\sim 15$\% may affect the polar-to-ecliptic ratio of the helioglow to an extent depending on the phase of the solar cycle. During solar minima, the ratio changes by 20--40\%, while during solar maxima it is rather 15--20\%. The stronger influence during solar minima seems to be related to the solar wind anisotropy, which amplifies the effects of the solar UV anisotropy. However, our further analysis of SOHO/SWAN observations shows that a strong solar wind anisotropy, typical for solar minima, and a strong solar UV anisotropy do not occur at the same time. For solar minima, the estimated solar UV anisotropy is rather weak, $\sim$5-10\%, and it may increase to $\sim$15-25\% during solar maxima.

The influence of the solar wind and solar UV anisotropies is also different in terms of polar brightening and darkening of the helioglow. The solar wind anisotropy increases the helioglow intensity in the polar sector and the solar UV anisotropy works typically in the opposite direction, i.e., decreases the polar intensity. These effects occur because the solar wind is faster and thinner at large latitudes, as revealed by Ulysses measurements \citep{mccomas_etal:03a, mccomas_etal:13b} and interplanetary-scintillation analysis \citep{tokumaru_etal:10a, sokol_etal:20a}. This results in lower ionization rates for hydrogen atoms, which moves the hydrogen cavity toward the Sun and increases the helioglow signal in polar regions. By contrast, the solar UV anisotropy is caused by the presence of polar coronal holes during solar minima and the clustering of the active regions at lower latitudes during solar maxima \citep{auchere_etal:05a}. The two causes of the UV anisotropy lead effectively to a polar darkening of the solar UV output, and thus also the helioglow, rather than polar brightening. From this point of view, a polar darkening of the helioglow can be interpreted rather as a signature of a significant solar UV anisotropy than related to the solar wind structure. Also, a latitudinally-flat helioglow profile cannot be attributed merely to the solar wind structure, since competition between the solar wind and the solar UV anisotropies may lead to some cancelation effects. This suggests that accounting for the solar UV anisotropy is important in inferring the solar wind structure from the Lyman-$\alpha$ helioglow observations.

Based on these findings, we derived the day-by-day time evolution of the inferred solar UV anisotropy separately for the northern and southern hemispheres from 1996 to 2022. In comparison with our previous work \citet{strumik_etal:21b}, this paper contains a more accurate estimation, where a better understanding of the anisotropy-related issues allowed us to address several difficulties. The refinements include better cleaning of observation artifacts, using time averages (to make observations and simulations more relevant to each other), and better normalization (based on the mid-latitude regions in the sky as the least affected by the anisotropy). Tracking the day-by-day time series gives us better insight into the time evolution and possible errors in the estimation of the anisotropy, resulting from temporary misfits between the simulations and observations. For this reason, we consider the yearly-moving-average-based estimate presented in Figure \ref{fig:anis_yearly_avg_infer} as more reliable and accurate in comparison with that shown in Figure \ref{fig:anis_simple_infer}. The time evolution of the anisotropy presented in Figure \ref{fig:anis_yearly_avg_infer} is consistent with an anisotropy obtained from synoptic maps of the Sun for 30.4 nm by \citet{auchere_etal:05a} for the years 1996-2003. The anisotropy evolution presented in our paper is also qualitatively consistent with the so-called butterfly diagrams for sunspots (see, e.g., \citet{leussu_etal:17a}). These diagrams illustrate the evolution of sunspots (which are typically dark regions in the visible range but bright in the UV). According to the diagrams, the sunspots are almost not visible during solar minima, then they appear in mid-latitude regions, their distribution becomes wider and they migrate towards lower latitudes as the transition to the solar maximum progresses. This time evolution seems to correspond to the behavior of the inferred anisotropy derived in our paper.

The new estimate of the anisotropy $A$ (blue and orange lines in Figure \ref{fig:anis_yearly_avg_infer}) has typically slightly smaller values (indicating a higher anisotropy) than the former estimation by \citet{strumik_etal:21b} (black circles) based on observations from the upwind positions of the observer every year. Analysis of Figure \ref{fig:anis_simple_infer} suggests that for the years 1996--2011 this difference is related mainly to spurious annual variations of the inferred anisotropy that appear when the yearly-averaging procedure is not used. For the years 2011--2022, the reason for the difference is presumably related to different normalizations used in the two papers. \citet{strumik_etal:21b} used a normalization based on the average value over an ecliptic-equatorial belt, while in this paper we use an improved normalization based on averages over mid-latitude regions, which are less affected by the solar UV anisotropy as discussed in Sections \ref{sec:maps_cmp} and \ref{sec:cuts_cmp}. The most striking (but still quite small) difference between the old and new estimate is presumably the inferred solar UV anisotropy level during solar minima, which is now $A \simeq$0.92 in comparison with $A \simeq 1$ in the former estimate. In this regard, the new results seem to be more consistent with the solar disk observations by \citet{auchere_etal:05a}, where a weak but non-negligible anisotropy is observed in the solar minimum and attributed to the coronal holes.

\acknowledgments{This study was supported by \added{the} Polish National Science Center grant 2019/35/B/ST9/01241 and by \added{the} Polish Ministry for Education and Science under contract MEiN/2021/2/DIR. We acknowledge the publicly available data with SOHO/SWAN satellite instrument observations provided by the following webpage http://swan.projet.latmos.ipsl.fr/data/L2/FSKMAPS. SOHO is a mission of international cooperation between ESA and NASA.}

\software{astropy \citep{astropy:13,astropy:18}}

\appendix
\section{SOHO/SWAN data cleaning}
\label{sec:appendix1}
\begin{figure}[!htbp]
	\begin{center}
		\includegraphics[scale=0.5]{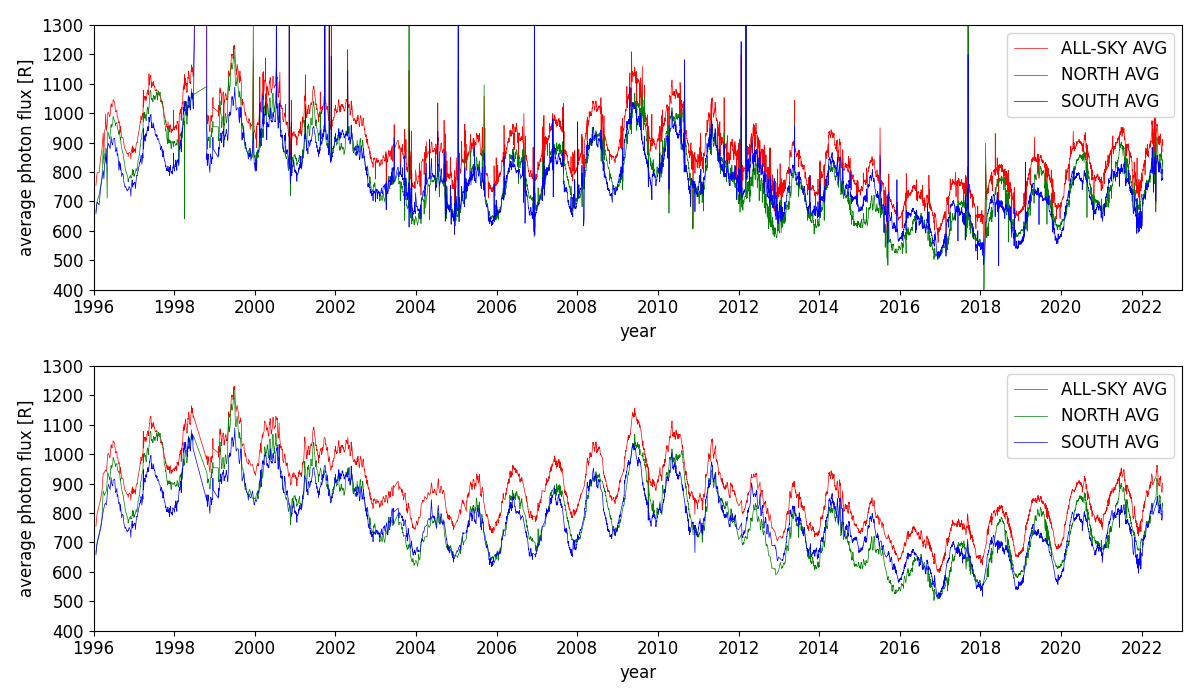}
		\caption{Comparison of averaged SOHO/SWAN observations before (top panel) and after (bottom panel) data-cleaning procedure used in this paper. Three types of averages are considered: all-sky (red line), northern hemisphere (green line), and southern hemisphere (blue line) in ecliptic coordinates. The hemispheric averages were computed after excluding the galactic equatorial belt ($\pm 15\degr$).}
		\label{fig:cleaning}
	\end{center}
\end{figure}

The SOHO/SWAN observations for particular dates may contain contaminations caused by solar events or various technical issues. For this reason, SOHO/SWAN observations used in this paper are selected from all available observation days to represent a regular evolution of the helioglow on time scales larger than the Carrington rotation period. Several criteria were used to exclude particular problematic dates. The first criterion excluded dates, for which masked regions in the sky (where observations are lacking) cover more than 15\% of the entire daily map. The second criterion excluded days, for which the average intensity (over the entire map) changed more than 30 R relative to the previous observation day. If a multi-day gap in the observations occurred extending over $N_\mathrm{D}$ days, this criterion assumed that $N_\mathrm{D}\times 30$ R is the maximum allowed variation of the average intensity between the previous and next available observation days, otherwise, the next observation day was excluded. We also excluded those observation days, for which MEDCPS26 field in the header of FITS files with the observation data was larger than 20 indicating possible solar-events contamination. By using these criteria, a majority of spikes were removed in the time series of the helioglow intensity evolution as seen in Figure \ref{fig:cleaning}. 
In this figure, the time evolutions of different types of averages are shown to illustrate the cleaning process described in this section. The time series in the bottom panel show solar-cycle ($\sim$11 years), one-year, and Carrington-rotation-period ($\sim$27 days) modulations. %The Carrington-rotation-period modulation is still present here, because the monthly-averaging procedure has not been applied at this stage yet.

\section{Normalization of helioglow intensities}
\label{sec:appendix2}
Sensitivity-degradation issues discussed recently by \citet{baliukin_etal:19a} and \citet{koutroumpa_etal:19a} for SOHO/SWAN can be expected to change the instrument calibration over time. These findings are confirmed by our comparison of the observed and simulated helioglow intensities discussed below.
\begin{figure}[!htbp]
	\begin{center}
		\includegraphics[scale=0.6]{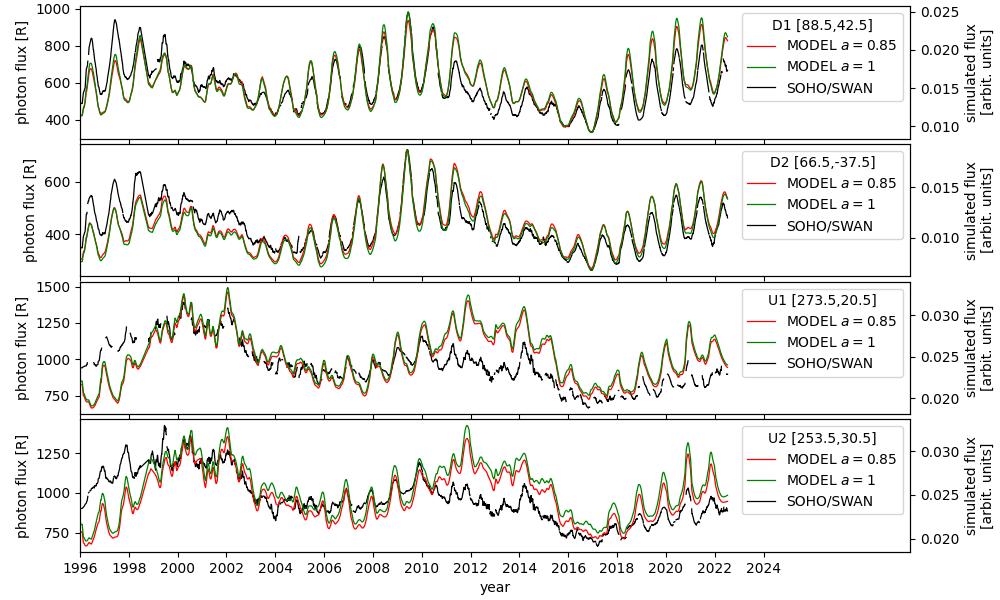}
		\caption{Comparison of WawHelioGlow simulations (red lines for the anisotropy $a=0.85$ and green lines for $a=1$) and SOHO/SWAN observations (black lines). The left vertical-axis label refers to SOHO/SWAN observations, while the right label -- to WawHelioGlow simulations. Two top panels show the time evolution of the raw (not-normalized) intensity for the downwind sector (directions D1 and D2), and the bottom two panels -- for the upwind sector (directions U1 and U2).}
		\label{fig:model_obs_cmp_not_norm}
	\end{center}
\end{figure}
Figure \ref{fig:model_obs_cmp_not_norm} shows the comparison of simulations and observations for the directions D1, D2, U1, and U2. The intensities are computed as 30-day moving averages to suppress variations that are usually present at the Carrington rotation time scale. Figure \ref{fig:model_obs_cmp_not_norm} shows that initially (1996-2000) the SOHO/SWAN flux (black line) is systematically higher than the simulated fluxes (red and green line), while later it becomes systematically smaller.

Following \citet{baliukin_etal:19a} and \citet{koutroumpa_etal:19a}, we attribute the systematic changes to calibration uncertainties and we decide to use normalized quantities for our comparisons of the simulated and observed fluxes. The normalization procedure is performed for two series of simulations, with the isotropic solar UV output ($\mathrm{SW3D/UV2D}$, where $a=1$) and with the anisotropic UV output ($\mathrm{SW3D/UV3D}$, where $a=0.85$), as well as for the SOHO/SWAN observations. For all of them, the normalized intensities were computed as $I(t) = I^\mathrm{raw}(t)/I^\mathrm{raw}_0(t)$, where $I^\mathrm{raw}(t)$ is the raw intensity (i.e., the observed and simulated photon flux in Figure \ref{fig:model_obs_cmp_not_norm}) in the respective simulation or observation units. The time-dependent normalization factor $I^\mathrm{raw}_0(t)$ is intended to have the capability of accounting for possible changes in the instrument calibration in time.

The factor $I^\mathrm{raw}_0(t)$ was computed in the following way. For the time evolution of the helioglow intensity from 1996 to 2022 (shown in Figure \ref{fig:model_obs_cmp_not_norm}), a yearly moving average $\overline{I^\mathrm{raw}(t)}$ was computed for directions D1, D2, U1, and U2. As showed in Figure \ref{fig:cmp_maps}(b) and (e), the directions are located at mid latitudes, where the difference between the models $\mathrm{SW3D/UV2D}$ and $\mathrm{SW3D/UV3D}$ is close to zero. In other words, these directions are weakly affected by the solar UV anisotropy. Therefore, this subset of points is suitable for computing the normalization factor $I^\mathrm{raw}_0(t)$, if anisotropy-related questions are of interest. The moving averages for the four points were then averaged to obtain one normalization factor $I^\mathrm{raw}_0(t) = [\overline{I^\mathrm{raw}_{D1}(t)}+\overline{I^\mathrm{raw}_{D2}(t)}+\overline{I^\mathrm{raw}_{U1}(t)}+\overline{I^\mathrm{raw}_{U2}(t)}]/4$. This procedure was applied to both simulation sets and observations separately. Although only four directions in the sky are used for defining the normalization, the yearly averaging procedure provides a significant rigidity and robustness to the normalization factor. Note also that for a given time, the normalization factor is applied globally over the entire sky (i.e., the same value is used for all directions D1, D2, U1, and U2).

Figure \ref{fig:model_obs_cmp} shows a similar comparison like Figure \ref{fig:model_obs_cmp_not_norm}, but for the normalized fluxes. The systematic changes of the observed fluxes relative to the simulated fluxes seen in Figure \ref{fig:model_obs_cmp_not_norm} are not present for the normalized quantities in Figure \ref{fig:model_obs_cmp}. This suggests, that systematic calibration changes of the observed intensity (on the time scales larger than one year) are naturally addressed by the procedure of normalization described above. 

%\bibliographystyle{aasjournal}
%\bibliography{iplbib}

\end{document}